\documentclass{emulateapj}

\catcode`\@=11
\newcommand{\gapprox}{\mathrel{\mathpalette\@versim>}}
\newcommand{\lapprox}{\mathrel{\mathpalette\@versim<}}
\newcommand{\propapprox}{\mathrel{\mathpalette\@versim\propto}}
\newcommand{\@versim}[2]
  {\lower3.1truept\vbox{\baselineskip0pt\lineskip0.5truept
\ialign{$\m@th#1\hfil##\hfil$\crcr#2\crcr\sim\crcr}}}
\catcode`\@=12
\journalinfo{}
\submitted{Accepted for publication in ApJ Letters on May 22, 2009}
\shorttitle{CSM SHELLS AROUND TYPE Ia SNe}
\shortauthors{BORKOWSKI, BLONDIN, \& REYNOLDS}

\begin{document}

\title{Circumstellar Shells in Absorption in Type Ia Supernovae}

\author{Kazimierz J. Borkowski,\altaffilmark{1}
John M. Blondin,\altaffilmark{1}
\& Stephen P. Reynolds\altaffilmark{1}
}

\altaffiltext{1}{Physics Department, North Carolina State U.,
    Raleigh, NC 27695-8202; kborkow@ncsu.edu}

\begin{abstract}
Progenitors of Type Ia supernovae (SNe) have been predicted to modify
their ambient circumstellar (CSM) and interstellar environments
through the action of their powerful winds. While there is X-ray and
optical evidence for circumstellar interaction in several remnants of
Type Ia SNe, widespread evidence for such interaction in Type Ia SNe
themselves has been lacking.
We consider prospects for the detection of CSM shells that have been
predicted to be common around Type Ia SNe.  Such shells are most
easily detected in \ion{Na}{1} absorption lines. Variable (declining)
absorption is expected to occur soon after the explosion, primarily
during the SN rise time, for shells located within $\sim$ 1--10 pc of
a SN. The distance of the shell from the SN can be determined by
measuring the time scale for line variability.
\end{abstract}


\section{CSM Shells around Type Ia Progenitors}
\label{type1acsm}

Type Ia supernovae (SNe) play an important role in many areas of
astrophysics, even though their progenitor systems and detailed
explosion mechanisms are poorly understood.  One fundamental problem
involves the prediction from many models, but so far nondetection, of
substantial circumstellar material (CSM), likely in the form of shells
at pc-scale distances.  Here, we predict the
decrease in \ion{Na}{1} absorption at early times that such shells 
should produce, a companion effect to Na absorption increases seen
somewhat later \citep{patat07}.  

Most popular current Type Ia scenarios involve single degenerate
(``SD'') progenitors, where the donor star supplies mass to the white
dwarf (WD), driving it close to the Chandrasekhar limit, where
instabilities produce an explosion.  A crucial step in this process is
the ability of an accreting white dwarf to expel material accumulating
on its surface in excess of a critical accretion rate through a fast
($v_w \sim 1000$ km s$^{-1}$) radiatively driven stellar outflow
\citep{hachisu96,hachisu99a,hachisu99b}.  In the absence of such a
wind, formation of a common envelope around the progenitor system is
unavoidable because of the expansion of the WD envelope. The companion
then merges with the white dwarf, preventing an explosion.

The WD accretion winds are expected to have profound consequences on
the CSM and interstellar medium (ISM) around Type Ia SNe.  Fast and 
powerful winds not only
sweep denser CSM out of the progenitor system, they have also been
suggested to strip material from the donor star 
\citep[e.g.,][]{hachisu08}
which then could form
an asymmetric disk in the binary orbital plane.
When swept by the SN blast wave, this material would be expected to
produce high velocity absorption features, and perhaps detectable
radio and X-ray emission.  Farther out, the white dwarf wind blows up
large-scale bubbles in the ambient ISM. \citet{badenes07} investigated
the evolution of these bubbles with one-dimensional hydrodynamical simulations.
For an ISM hydrogen density $n_0$ of 0.43 cm$^{-3}$ ($\rho_0 =
10^{-24}$ g cm$^{-3}$) and $v_w = 10^3$ km s$^{-1}$, large ($> 15$ pc)
bubbles are blown by progenitor winds, sweeping the ambient ISM into
massive ($> 300$ $M_\odot$) shells.

Since typical progenitor space velocities $v_*$ of $\sim 20$ km
s$^{-1}$ are larger than final shell velocities, a SN explosion
generally occurs after the progenitor has moved away from the bubble's
center, by a distance depending on $n_0$ and on the progenitor
scenario under consideration. If the WD explodes early ($\sim 5 \times
10^5$ yr) after the onset of the WD wind
\citep[as in models H1 and HP3 of][]{badenes07}, the explosion occurs
off-center but still inside the bubble. For $n_0 = 0.43$ cm$^{-3}$ and 
$v_* = 20$ km s$^{-1}$, the bubble's radius is 17 (22) pc for model H1 (HP3),
while the SN exploded 7 (9) pc away from the shell.  The SN would be
deeper inside the bubble for slower moving progenitors and less dense
ambient ISM. For faster progenitors and denser ISM, an explosion could
even occur outside the bubble, but still in the vicinity of the
swept-up shell.

Stellar motions are most important for progenitors that explode long
after the onset and cessation of the wind \citep[$2 \times 10^6$ and
$10^6$ yr, respectively, in models HP1 and L2 of][]{badenes07}.  A
progenitor with $v_* = 20$ km s$^{-1}$ travels 40 pc in $2
\times 10^6$ yr, beyond bubbles with a radius of 27 (36) pc in model HP1
(L2) (for $n_0 = 0.43$ cm$^{-3}$). Without a fast wind blowing at the time of
the explosion, these SNe explode in the undisturbed ISM.

Another outcome is expected for a progenitor with a moderate,
long-duration wind that is still blowing when the SN explodes
\citep[as in models L1 and LV1 of][]{badenes07}.  Even if the
progenitor left the bubble's interior, the wind interaction with the
ambient ISM leads to the formation of a bow shock ahead of the moving
SN progenitor \citep[e.g.,][]{wilkin96}. The standoff distance $R_0$
of the swept-up ISM shell in the direction of the stellar motion is
0.75 pc for a mass-loss rate $\dot{M}_w = 10^{-7} M_\odot$ yr$^{-1}$,
$v_w \sim 10^3$ km s$^{-1}$, $v_* = 20$ km s$^{-1}$, and $n_0 = 1$
cm$^{-3}$ ($R_0 \propto \dot{M}_w^{1/2} v_w^{1/2} n_0^{-1/2} v_*^{-1}$). 
The distance to the shell is 1.3 pc in the plane
perpendicular to the stellar motion direction and passing through the
progenitor. The shell surface density is $\sim R_0 n_0$ 
\citep[$=2.3 \times 10^{18}$ cm$^{-2}$ for $R_0 = 0.75$ pc and $n_0 = 1$ 
cm$^{-3}$;][]{wilkin96,comeron98}. As discussed next, such a CSM shell and
more massive CSM shells swept up by the WD winds are expected to
produce detectable absorption lines in SN spectra.

CSM has also been revealed by young Type Ia 
supernova remnants
(SNRs). Kepler's SN was a Type Ia explosion within a dense CSM
\citep{reynolds07}. The total shocked CSM mass in Kepler's SNR is
$\sim 0.75 M_\odot$ \citep{blair07}. The blast wave propagated to 2 pc
away from the explosion center, and it is still encountering dense
material with a preshock density of $\sim 5$ cm$^{-3}$. This CSM was
ejected by the SN progenitor, unlike the massive swept-up ISM shells
just discussed. We generally expect shells at small radii to be
dominated by ejected CSM, with the swept-up ISM dominating at large
radii.

\section{A Simple CSM Shell Model}
\label{shellmodel}

A geometrically thin, spherically symmetric CSM shell with radius $R_{s}$ 
and mass $M_{s}$ has a hydrogen column density  
\begin{equation}
N(H) = 1.78 \times 10^{18} \left( \frac{R_{s}}{2\ {\rm pc}} \right)^{-2} 
\frac{M_{s}}{M_\odot}\ {\rm cm}^{-2}.
\label{hcolumn}
\end{equation}
For a shell swept up by a progenitor WD wind in 
a uniform ISM with density $n_0$, equation (\ref{hcolumn})
becomes
\begin{equation}
N(H) = 2.06 \times 10^{18} \frac{R_{s}}{2\ {\rm pc}} 
\frac{n_0}{{\rm cm}^{-3}} \ {\rm cm}^{-2}.
\label{hcolumnism}
\end{equation}
Theoretical considerations and observational evidence suggest a wide range
in shell masses, from a fraction of a solar mass at sub-pc scales to 
hundreds of solar masses at $\sim 10$ pc scales (Section \ref{type1acsm}). 
This simple CSM shell model does not encompass progenitors
that left the ambient ISM undisturbed. 

Shells with N(H) $\sim 10^{18}$--$10^{19}$ cm$^{-2}$
(eqs.~\ref{hcolumn} and \ref{hcolumnism}) can be most easily detected
in optical \ion{Na}{1} and \ion{Ca}{2} absorption lines. We focus on
\ion{Na}{1} lines, because of the likely depletion of Ca onto dust that
may reduce \ion{Ca}{2} absorption below detectable levels.  The
CSM dust found in Kepler \citep{blair07}, for example, would result in
significant depletion of Ca.
Neutral Na is quickly ionized by the Galactic UV radiation field
because of its low (5.14 eV) ionization potential. For the
``standard'' UV field \citep[e.g.,][]{peqald86}, the ionization rate
$\Gamma$(\ion{Na}{1}) is equal to $1.0 \times 10^{-11}$ s$^{-1}$.
As in the general ISM, the Na ionization state is determined by the
balance between ionizations and recombinations,
$n$(\ion{Na}{1})$\Gamma$(\ion{Na}{1}) $= n_e n$(\ion{Na}{2})
$\alpha_{Na II} (T_e)$, where $\alpha_{Na II}$ is the recombination
coefficient for \ion{Na}{2}.  For gas with predominantly singly
ionized Na with a solar (cosmic) abundance of $2.0 \times 10^{-6}$ 
\citep[by number with respect to H;][]{lodders09} and ``standard'' 
Galactic UV radiation
field, N(\ion{Na}{1})/N(H)$=2.2 \times 10^{-7} n_e \alpha_{{\rm Na
II}}(T_e)/\alpha_{{\rm Na II}}(10^3 K)$,
and 
\begin{eqnarray}
\lefteqn{N({\rm Na I}) = } \\  \nonumber
& & 3.9 \times 10^{10} 
\frac {\alpha_{{\rm Na II}}(T_e)}{\alpha_{{\rm Na II}}(10^3 K)}
\frac {n_e}{0.1~{\rm cm}^{-3}}
\left( \frac{R_{s}}{2\ {\rm pc}} \right)^{-2} 
\frac{M_{s}}{M_\odot}\ {\rm cm}^{-2}. 
\label{nacolumn}
\end{eqnarray}
For a shell swept up by the progenitor WD wind, 
\begin{eqnarray}
\lefteqn{N({\rm Na I}) = } \\ \nonumber
& & 
4.6 \times 10^{10} 
\frac {\alpha_{{\rm Na II}}(T_e)}{\alpha_{{\rm Na II}}(10^3 K)}
\frac {n_e}{0.1~{\rm cm}^{-3}}
\frac{R_{s}}{2\ {\rm pc}} 
\frac{n_0}{{\rm cm}^{-3}}\ {\rm cm}^{-2}.
\label{nacolumnu}
\end{eqnarray}
The electron density $n_e$ (and to a lesser degree also the electron
temperature $T_e$) within the CSM shell are additional parameters that
control the abundances of trace ions such as \ion{Na}{1} and
\ion{Ca}{2}.  Here, some degree of ionization of H produces the free
electrons that permit the existence of neutral Na.
A modest photon ionizing flux, provided by a SN progenitor or by the
ambient extreme UV Galactic radiation field, is likely to ionize a
small ($\sim 0.1$) fraction of H atoms, so the compressed shell gas 
at typical densities of $\sim 1$
cm$^{-3}$ \citep{badenes07} is expected to be detectable in \ion{Na}{1}
lines.  

In the SD progenitor scenario, ionizing photons are produced by the WD
and its accretion disk. For example, \citet{chugai08} considered a
blackbody with $T = 3 \times 10^4$ K and $L = 300 L_\odot$ as an
ionizing source in his considerations of
absorption lines in SNe with symbiotic progenitors.
Such a modest luminosity source is capable of significant ionization 
of the ambient gas to large distances away from the progenitor.
We demonstrate this by performing photoionization calculations with
Cloudy \citep[ver.~08.00;][]{ferland98}.  In addition to the
blackbody radiation produced by the progenitor, we also include the
diffuse Galactic UV radiation field \citep{draine96} that is
responsible for the ionization of \ion{Na}{1}.  A CSM shell with 1
$M_\odot$, $R_{s} = 2$ pc, and $n = 5$ cm$^{-3}$ becomes nearly
completely (99\%) ionized by the progenitor radiation.
But Na remains singly ionized, leading to an appreciable ($5 \times
10^{10}$ cm$^{-2}$) \ion{Na}{1} column density.
For a more massive (10 $M_\odot$) and distant (10 pc) shell, the H
ionization decreases to 80\%; N(\ion{Na}{1}) drops slightly to $4 \times
10^{10}$ cm$^{-2}$.
N(\ion{Na}{1}) increases with the increasing shell mass; for 50
$M_\odot$, it is equal to $2 \times 10^{11}$ cm$^{-2}$, while H
becomes partially (49\%) neutral.
There are large spatial variations of the diffuse
ionizing radiation in galaxies, and its penetration into the CSM shell
depends strongly on the assumed geometry and dynamical history of the
shell. In view of these large uncertainties, we simply use
equations~(3) and (4) in our estimates of
\ion{Na}{1} column densities, using $n_e$ as a free parameter to
describe the unknown ionization fraction of H in the CSM shell with
density $\sim 1$--10 cm$^{-3}$.

\section{Variable \ion{Na}{1} Absorption}
\label{absorption}

\begin{figure}
\plotone{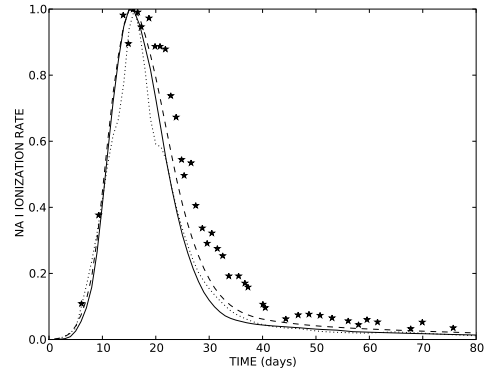}
\caption{Normalized \ion{Na}{1} ionization rate as a function of time since 
explosion (solid line: based on spectral templates of Nugent et
al.\,2002; dotted line: on the SNLS templates).  A synthetic
{\it Swift uvw2} light curve (dashed line; based on Nugent et
al.\,2002), and normalized {\it uvw2} fluxes of SN\,2007af 
(stars) are also plotted.
\label{na1ionization} }
\end{figure}

Photoionization of \ion{Na}{1} by SN photons is very efficient in the
immediate ($\le$ several $\times 10^{17}$ cm) vicinity of the SN
\citep{chugai08}, decreasing rapidly
with distance. 
We considered photoionization of \ion{Na}{1} as a function
of shell location $R_{s}$ and time $t$ since the explosion, using Type
Ia Branch-normal spectral templates
\citep{nugent02}\footnote{Available online at
www.supernova.lbl.gov/$\sim $nugent} and \ion{Na}{1} photoionization
cross sections from \citet{verner96}. Only photons with energies
larger than the \ion{Na}{1} ionization potential of 5.14 eV ($\lambda
< 2410$\AA) are capable of ionizing neutral sodium. 
Figure \ref{na1ionization} shows the
resulting \ion{Na}{1} ionization rate as a function of time,
normalized to 1 at its maximum.  With $M_V=-19.46$ at maximum light
\citep{tammann08}, the normalizing factor is equal to $3.3 \times
10^{-6} \left( R_{s}/2\ {\rm pc} \right)^{-2}$ s$^{-1}$.  We also show
in Figure \ref{na1ionization} the normalized \ion{Na}{1} ionization
rate based on spectral templates from the Supernova Legacy Survey
\citep[SNLS;][]{hsiao07}\footnote{Available online at
www.astro.uvic.ca/$\sim $hsiao}. The normalizing factor is only 7\%\
smaller than for Nugent's templates.
Flux measurements with the {\it Swift} satellite's {\it uvw2} filter
($\lambda_{central}=1928$\AA, FWHM$=657$\AA) are of particular
interest, since they provide light curves shortward of the \ion{Na}{1}
ionization limit. Aside from a noticeable excess at late times caused
by a red leak in this filter, a synthetic {\it uvw2} light curve
matches the \ion{Na}{1} photoionization rate well
(Figure \ref{na1ionization}). At its maximum on day 15, $M_{uvw2} =
-17\fm5$ \citep[using {\it Swift} UVOT calibration by][]{poole08}.
For comparison, we also plot a normalized {\it uvw2} light curve for
SN\,2007af \citep[from][]{brown09}, a nearby, well-observed normal 
Type Ia SN. We assumed a
typical Type Ia rising time of 18 days \citep{garg07} in the {\it B} band
for SN\,2007af.  The calculated \ion{Na}{1} ionization rate curve is
narrower than the {\it uvw2} light curve of SN\,2007af by about 3
days, while the total number of \ion{Na}{1} ionizing photons is likely
to be reasonably well modeled by the template (based on a dereddened
$M_{uvw2}$ of $-16\fm8$ for this SN).
This modest mismatch likely reflects systematic deficiencies in our
current knowledge of Type Ia UV spectra, due to their UV
faintness, uncertain reddening corrections, and intrinsic UV
spectral variations.

\begin{figure}
\plotone{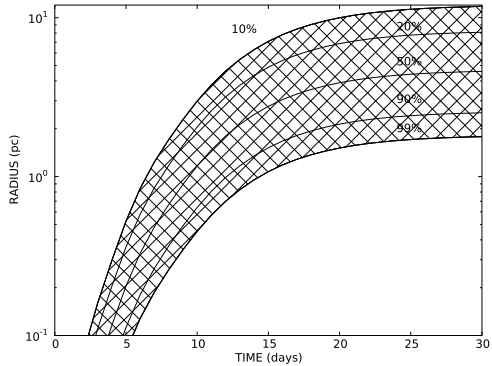}
\caption{Decrease in \ion{Na}{1} absorption as a function of time and
distance from the SN. The curves are labeled by the amount of decrease in
N(\ion{Na}{1}).  Reasonable prospects for the detection of absorption line
variability occur within the cross-hatched area.
\label{na1variability} }
\end{figure}

Figure \ref{na1variability} shows how \ion{Na}{1} absorption decreases
as a function of $R_{s}$ and $t$, using the spectral templates of
\citet{nugent02}.  Because the \ion{Na}{1} ionization rate peaks just
before the {\it B-}band maximum at $\sim 18$ days
(Figure \ref{na1ionization}), variable absorption is expected in the
first 20--30 days after the explosion, primarily during the SN rise
time.  Detection will be difficult for decreases below 10\% 
or where 99\% of Na ions are ionized.
Prospects for detection of variability are optimal in the middle of
the cross-hatched area in Figure \ref{na1variability}, and
observational constraints favor detection of shells at pc-scale
distances.

\begin{figure}
\plotone{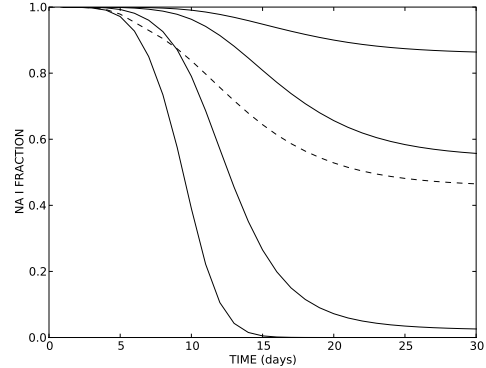}
\caption{Remaining \ion{Na}{1} fraction as a function of time since explosion 
for shells located at 1, 2, 5, and 10 pc (solid lines from 
left to right) from the SN. The dashed curve corresponding to the uniform 
ambient ISM is also shown. 
\label{na1fraction} }
\end{figure}

Figure \ref{na1fraction} shows the remaining \ion{Na}{1} fraction for
shells with $R_{s} = 1$, 2, 5, and 10 pc. These curves depict
variations in N(\ion{Na}{1}) (or the equivalent width of the Na D
lines in the limit of low line optical depths) as a function of
time. The rate of change in absorption is much faster for shells at
smaller $R_s$, reflecting distance-dependent ($\propto R_s^{-2}$)
variations in the Na ionization rate. This sensitivity to distance
from the SN can be utilized to infer distances of CSM shells from
observations of absorption line variability.  We also show in Figure
\ref{na1fraction} the remaining \ion{Na}{1} fraction for uniformly (up
to 10 pc) distributed ISM. Except at early times, variations in
absorption are rather similar to variations produced by the shell with
$R_{s}=5$ pc. 
But absorption lines
from an undisturbed ISM will be harder to detect than from the CSM
shells, because these lines are more likely to be located within
saturated cores of the \ion{Na}{1} D absorption line profiles produced
by the ISM in host galaxies.

Intrinsic variations in the UV among SNe are likely to affect the
detectability of \ion{Na}{1} lines. Overluminous SNe such as SN\,1991T
are attractive targets, because their higher than average fluxes
facilitate high spectral resolution observations of absorption lines,
while their potentially higher UV luminosities may allow probing of
the swept-up ISM to larger distances. The brightest SNe are
also associated with star-forming galaxies \citep[e.g.,][]{sullivan06},
implying more gas-rich environments near overluminous
SNe. Conversely, subluminous SNe such as SN\,1991bg are more difficult
to observe, and their potentially lower UV luminosities might allow 
probing of the CSM only to sub-pc distances from the SN.

\section{Discussion}

Variable \ion{Na}{1} absorption in Type Ia SN spectra probes CSM at
pc-scale distances from the SN, such as the CSM seen in Kepler's SNR.
With $R_{s} = 2$ pc, $M_{s} = 0.75 M_\odot$, 
and $n_e = 0.5$ cm$^{-3}$, the preexplosion N(\ion{Na}{1}) is $1.5
\times 10^{11}$ cm$^{-2}$ (from eq.~3 with $T_e = 10^3$
K). The decrease in absorption is shown in Figure \ref{na1fraction}
(second curve from left); N(\ion{Na}{1}) drops below $10^{10}$
cm$^{-2}$ on day 20. Sufficiently rapid motion 
\citep[radial velocity of $-230$ km s$^{-1}$;][]{blair91} 
of Kepler's SN progenitor would have assured no
overlap with the (presumably) saturated ISM absorption in the line of
sight to Kepler's SN. It is possible to measure the predicted
\ion{Na}{1} D absorption in Type Ia spectra at early times.
\citet{patat07b} obtained a stringent ($2 \times 10^{10}$ cm$^{-2}$) 
upper limit for SN\,2000cx.
\citet{simon07} observed SN\,2007af at three epochs (days 14,
35, and 42 in Figure \ref{na1ionization}).  Their observation on day 14
was the least restrictive; the 5$\sigma$ limit on the CSM absorption
is $9 \times 10^{10}$ cm$^{-2}$. The predicted column density on day
14 is $5 \times 10^{10}$ cm$^{-2}$, at the detection limits of this
observation. No detection would have been expected on days 35 and
42. Our simple shell model for Kepler is of course uncertain. Kepler's
SNR is very asymmetric; depending on viewing angle, much lower
or much higher column densities are possible in a SN with such
strongly asymmetric CSM.  Kepler's SN also has a central emission
enhancement, indicating the presence of CSM at distances less than 2
pc from the SN; that would result in faster ionization by the SN.

The remnant of SN 185, SNR RCW 86, probably resulted from a Type
Ia explosion within a bubble \citep{badenes07}. At $\sim$2.5 kpc 
away \citep{sollerman03},
its 21$'$ radius corresponds to 15 pc. Assuming a
spherically symmetric bubble with this radius, the mass of the
swept-up ISM is equal to $500 n_0$ M$_\odot$. From Equation
(4), the preexplosion N(\ion{Na}{1}) is larger than
$10^{11}$ cm$^{-2}$ for $n_0 > 0.29$ cm$^{-3}$ (with $n_e = 0.1$
cm$^{-3}$ and $T_e = 10^3$ K). RCW 86 is much brighter in the SW than
elsewhere, suggesting an off-center explosion in this region of the
SNR. The offset of the SN progenitor from the bubble's center is not
known; we assume $\sim 10$ pc offset from the center ($\sim 5$ pc
distance from the shell). The decrease in absorption is shown in Figure
\ref{na1fraction} (second curve from top); half the Na atoms are
ionized in the shell section closest to the SN. Significant ($> 10$\%)
variations in N(\ion{Na}{1}) would be seen by about half of
randomly distributed observers; for the other half, the CSM shell
would have been too far away from the SN to produce variable
absorption.  The predicted column densities depend sensitively on 
geometric details, but our simple model for RCW 86
demonstrates that shells swept by SN progenitors should be detectable
in \ion{Na}{1} lines.

The variable \ion{Na}{1} absorption detected in SN\,2006X
\citep{patat07} and SN\,1999cl \citep{blondin09} cannot be due to our 
CSM shell effect, as it is stronger, slower, and increasing rather than
decreasing.
While the origin of these variations is still not understood,
both SNe are among the most highly reddened Type Ia SNe, leading
\citet{blondin09} to suggest a connection to dusty environments in
their host galaxies. We expect a correlation between variable
absorption and the reddening within host galaxies, because formation
of massive wind-swept shells at pc-scale distances from the SN can
occur only in gas-rich hosts. But the line variability is confined to
early times (Figure \ref{na1fraction}), and the predicted decline 
is obviously in conflict with the much larger 
increase in N(\ion{Na}{1}) observed in SNe\,2006X and 1999cl.   

Typical velocities of shells swept up by the progenitor WD winds are 
expected to be fairly low, $\sim 10$ km s$^{-1}$, comparable to the 
intrinsic velocity dispersion of the diffuse ISM, because 
the SN explosion usually occurs in late stages of bubble evolution 
after the shell has been effectively decelerated by the ambient ISM. 
When the shell radial velocity falls within the saturated or 
partially saturated \ion{Na}{1} D line profile of the host galaxy, 
prospects for the shell detection are markedly reduced. This strongly 
favors detection of blueshifted line absorption preferentially 
produced by progenitors with negative radial velocities (in the host 
galaxy frame), comprising no more than 50\%\ of all SNe. Another 
important selection effect is associated with the ambient medium density; 
formation of massive shells cannot occur in the low-density, hot ISM 
phase with a typical volume fraction of $\sim 0.5$. As a result, 
variable absorption is expected in only a fraction (certainly no more 
than a quarter) of all Type Ia SNe. 

Detection of variable \ion{Na}{1} absorption from CSM shells is more
difficult than for SNe\,2006X and 1999cl, and requires high-resolution
spectroscopy soon after a SN discovery. But such observations can map
the ambient medium around many Type Ia SNe, allowing for the detection
of CSM shells and establishing their distances from the SN by
measuring the rate of decrease in \ion{Na}{1} absorption. Most
progenitor scenarios predict the formation of shells at pc-scale
distances in many progenitor systems. This prediction can be verified
by observations of variable absorption, advancing our knowledge of
poorly understood Type Ia progenitors. We can learn much about these
progenitors by correlating variable absorption with SN properties such
as luminosities, UV excesses, or the presence or absence of
high-velocity features. Observational efforts focused on high spectral
resolution observations of Type Ia SNe following soon after their
discovery are warranted.
\acknowledgments 
We thank the referee, Nikolai Chugai, for fruitful comments on our 
original manuscript. This work was supported by NSF grant AST-0708224. 


\end{document}